\documentclass{PoS}

\title{Gross-Neveu model as a laboratory for fermion discretization}

\ShortTitle{Gross-Neveu model}

\author{Tomasz Korzec and \speaker{Ulli Wolff}\\
        Humboldt University, Newtonstr. 15, 12489 Berlin, Germany\\
        E-mail: \email{uwolff@physik.hu-berlin.de},
        \email{korzec@physik.hu-berlin.de}}

\abstract{
We introduce a finite volume renormalization scheme
for the $N$-Majorana-component O($N$) invariant  Gross-Neveu model.
Universal observables are defined that are accessible
to precise numerical simulation in various discretizations
and allow for an extrapolation to the continuum limit.
Here first numerical results with Wilson fermions are reported.
For $N=2$ they reproduce exact finite volume continuum results 
in the massless Thirring
model. Our $N=8$ data are ready for comparison for instance with staggered
results in the future.
\vspace{3cm}
\begin{flushright}
HU-EP-06/29\\
SFB/CPP-06-44
\end{flushright}
}

\FullConference{XXIVth International Symposium on Lattice Field Theory\\
                July 23-28, 2006\\
                Tucson, Arizona, USA}

\newcommand{\be}{\begin{equation}}
\newcommand{\ee}{\end{equation}}
\newcommand{\bes}{\begin{eqnarray}}
\newcommand{\ees}{\end{eqnarray}}

\newcommand{\rO}{\mathrm{ O}}
\newcommand{\mathe}{\mathrm{e}}

\begin{document}

\section{Introduction}
In lattice QCD several fermion discretizations are in use in current
dynamical simulations. Beside the very costly recent variants with chiral symmetry,
Wilson and staggered fermions are the standard choices with their well-known
relative merits and weaknesses. For the latter choice, unphysical multiples
of 4 degenerate flavors (in 4 dimensions) can only be avoided by the `rooting'
procedure which is under much debate \cite{SS}, as universality of the continuum limit
is not guaranteed any more by the locality of the action.

We thought that in this situation a two-dimensional study, where the continuum
extrapolation can be controlled better than in QCD, would be a valuable check.
Gross-Neveu models (GN) \cite{Gross:1974jv}  
and the Schwinger-model come to mind. The latter is closer to
QCD in being a gauge theory, while the renormalization structure of GN
is more realistic as its coupling is dimensionless, even asymptotically free
for $N \ge 3$, instead of being superrenormalizable. For us the latter aspect
prevails.

The work of the ALPHA collaboration has demonstrated, that a precise continuum
extrapolation becomes feasible for quantities like the Schr\"odinger functional
of QCD, where the system size is used as a physical scale to probe the field
theory. We hence construct a similar finite volume renormalization scheme for
GN. Also here
the finite size supplies an infrared scale allowing the mass to be tuned
to a critical value\footnote{It vanishes if the chiral symmetry is preserved
by the regularization.}
corresponding to the (here discrete) chirally 
symmetric continuum limit.
With the coupling as the only remaining free parameter, the situation becomes
similar to the massless QCD Schr\"odinger functional. While the aim clearly
is to use this scheme also for staggered fermions, possibly with `rooting',
in the future, at present we only have Wilson simulations to report on.

\section{Model and renormalization scheme}

We consider the action density of the Euclidean theory
\be
\mathcal{L} =  \frac{1}{2} \xi^{\top}
   \mathcal{C}( \not\partial + m )
   \xi
-
\frac{g^2}{8} ( \xi^{\top}
   \mathcal{C} \xi )^2.
\label{GNaction}
\ee
Here the Grassmann spinor $\xi$ caries a two-valued spin and an $N$-valued
flavor index such that an internal O($N$) symmetry arises and 
the antisymmetric matrix $\mathcal{C}$
obeys $\mathcal{C}\gamma_{\mu}\mathcal{C}^{- 1} 
= - \gamma_{\mu}^{\top}$.
With this symmetry no other 4-fermion interaction is possible
and the model is renormalizable as it stands. This is in contrast
to the chiral GN model with $N$ Dirac fields, which shares continuous
chiral symmetry with QCD, but allows for two independent couplings
and a third one with Wilson fermions \cite{Korzec:2005ed}. This
makes it much more difficult to approach a definite continuum theory
and we hence restrict ourselves for our fermion-testbed to the O($N$) invariant class.
For $m=0$ the action (\ref{GNaction}) has the additional discrete
invariance
\be
\xi \to \gamma_5 \, \xi
\label{chisym}
\ee
that breaks spontaneously for $N=\infty$ \cite{Gross:1974jv}.

Our action (\ref{GNaction}) has so far referred to the continuum.
In the Wilson $r=1$ lattice regularizaton we have to replace
\be
\partial_\mu \to \tilde{\partial}_\mu, \quad 
m \to m - \frac{a}{2} \partial_\mu^{\phantom{\ast}} \partial_\mu^\ast 
\label{Wilsonterm}
\ee
with the forward ($\partial$), backward ($\partial^\ast$) 
and symmetric ($\tilde{\partial}$) lattice derivative.
Now (\ref{chisym}) is not a symmetry at finite cutoff. It emerges however
in the continuum limit for a suitable tuning $m=m_c(g^2)$.

For our finite volume scheme we consider the field $\xi$ on a $T\times L$ torus
with (anti)periodic boundary conditions in space (time)
\be
\xi(x+T\hat{0})=-\xi(x), \quad \xi(x+L\hat{1})=+\xi(x),
\ee
i.~e. a spatial ring at finite temperature. In the following we take $T=L$,
aspect ratio one, and
the smallest momentum then is $p_\ast=(\pi/T,0)$. In a momentum version of our 
finite size
scheme we formulate a complete set of normalization conditions 
on 2- and 4-point functions using external momenta $\pm  p_\ast$. We however
here prefer
to use correlations at physical separations in space-time for numerical reasons
and also in the hope -- supported by perturbation theory --  
to minimize cutoff effects in this way.

We Fourier-transform in space only
\be
\breve{\xi}(x_0,p)=a \sum_{x_1} \mathe^{-ipx_1}\xi(x),\quad 
\breve{\xi}(x_0)\equiv\breve{\xi}(x_0,0)
\ee
and impose normalizations on the 2-point function
\bes
0 &=&
\left\langle 
\breve{\xi}^\top (T/4) \mathcal{C} \breve{\xi}(0)
\right\rangle, \label{mccond}\\
Z_\xi &=& \frac{-1}{NL}
\left\langle 
\breve{\xi}^\top (T/2) \mathcal{C}\gamma_0 \breve{\xi}(0)
\right\rangle.
\ees
Now $\xi_R=Z_\xi^{-1/2}\xi$ is a renormalized field, and the first condition
is required by (\ref{chisym}) and determines the critical mass $m_c$.
Note that at separation $T/2$ the analogous equation holds as a consequence
of antiperiodicity and time-reflection invariance for all $m$ and could not serve to
define $m_c$, while with our choice we found good sensitivity to do so.
On the lattice only $T/a$ that are multiples of 4 must be simulated
to obtain a scaling situation, an acceptable restriction.
Finally a renormalized coupling is obtained from
\be
g_R^2 = \frac{4}{TL}  
\left\langle 
\breve{\xi}_{R,1}^{\top} ( T / 2 )\mathcal{C} \breve{\xi}_{R, 1} ( 0 ) \;
\breve{\xi}_{R,2}^{\top} ( T / 2 )\mathcal{C} \breve{\xi}_{R, 2} ( 0 ) 
\right\rangle,
\label{gRdef}
\ee
where the subscripts on $\xi_R$ refer to two specific flavor values.
For Wilson fermions we have performed a 1-loop calculation and obtain
\bes
a m_c &=& - ( N - 1 ) K g^2 + \rO ( a^4 g^2, g^4 ),
   \quad K = 0.384900179460 \\
Z_\xi &=& 1+\rO(g^4)\\
g_R^2 &=& \frac{T}{T - 2 a}  \left( g^2 + \left[
   (N - 2)\left(\frac{\ln ( L / a )}{2 \pi} + c_0\right)+c_1 + \rO ( a ) \right]
   g^4 + \rO ( g^6 ) \right) \label{gR1loop}
\ees
with
\be
c_0=-0.483524477,\quad c_1 = 0.30965176.
\ee
The 1-loop value for $m_c$ agrees with the literature 
\cite{Aoki:1985jj,Kenna:2001fs,Leder:2005mm}.
The coefficient of the logarithm in (\ref{gR1loop}) has the well known
value and vanishes for the Thirring case $N=2$. Note that
a finite coupling renormalization as well as the
linearly divergent mass renormalization are still there.

\section{Simulation}
The standard approach for GN is to factorize the 4-fermion term with a
scalar auxiliary field
\be
\mathe^{\frac{g^2}{8} a^2 \sum_x ( \xi^{\top}\mathcal{C} \xi )^2} 
= \int \prod_x d\mu ( \sigma ) 
  \mathe^{- \frac{g}{2} a^2 \sum_x \sigma \xi^{\top}
  \mathcal{C} \xi}.
\ee
Usually a Gaussian field is employed giving
\footnote{
At finite $N$ only a finite number of moments in $\sigma$
are relevant allowing in principle also other distributions
}
\be
Z = \int \prod_x d \sigma \,
\mathe^{-\frac{1}{2} a^2 \sum_x \sigma^2} [{\rm Pf}\, A ( \sigma )]^N,
\ee
where the Pfaffian results from integrating out $\xi$.
The operator
\be
 A ( \sigma ) =\mathcal{C}( \gamma_{\mu} \tilde{\partial}_{\mu} + m + g \sigma - a
   \partial^{\ast} \partial )
\ee
can be taken real antisymmetric in the Majorana representation of $\gamma_\mu$.
Hence for even $N$ we have the non-negative weight 
$0\le [{\rm Pf}\, A]^N=[{\rm det A^\top A}]^{N/4}$ which may
be represented by $N/2$ {\em real} pseudofermions for an HMC approach.

\section{Massless Thirring model ($N=2$)}

\begin{figure}[ht]
\centering
\includegraphics[width=.6\textwidth]{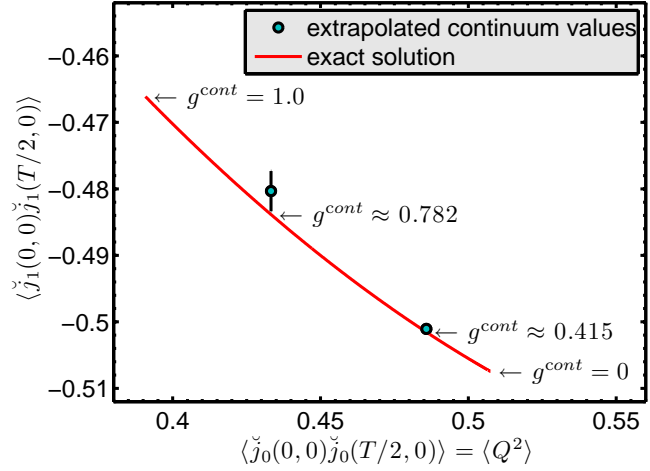}
\caption{Exact universal relation between current correlations in the
massless Thirring model ($N=2$).}
\label{jjfig}
\end{figure}
\begin{figure}[ht]
\centering
\includegraphics[width=.45\textwidth]{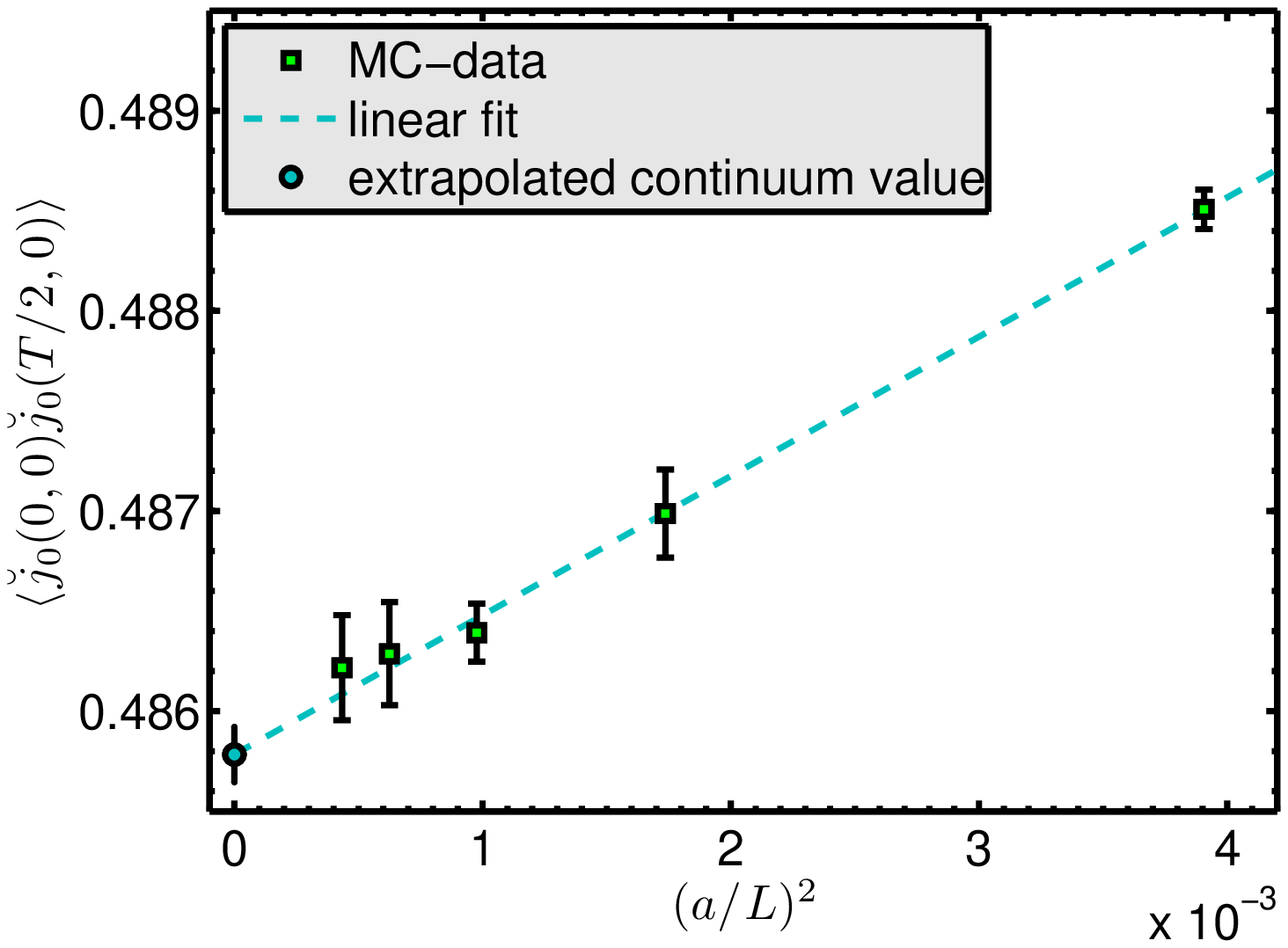}
\includegraphics[width=.45\textwidth]{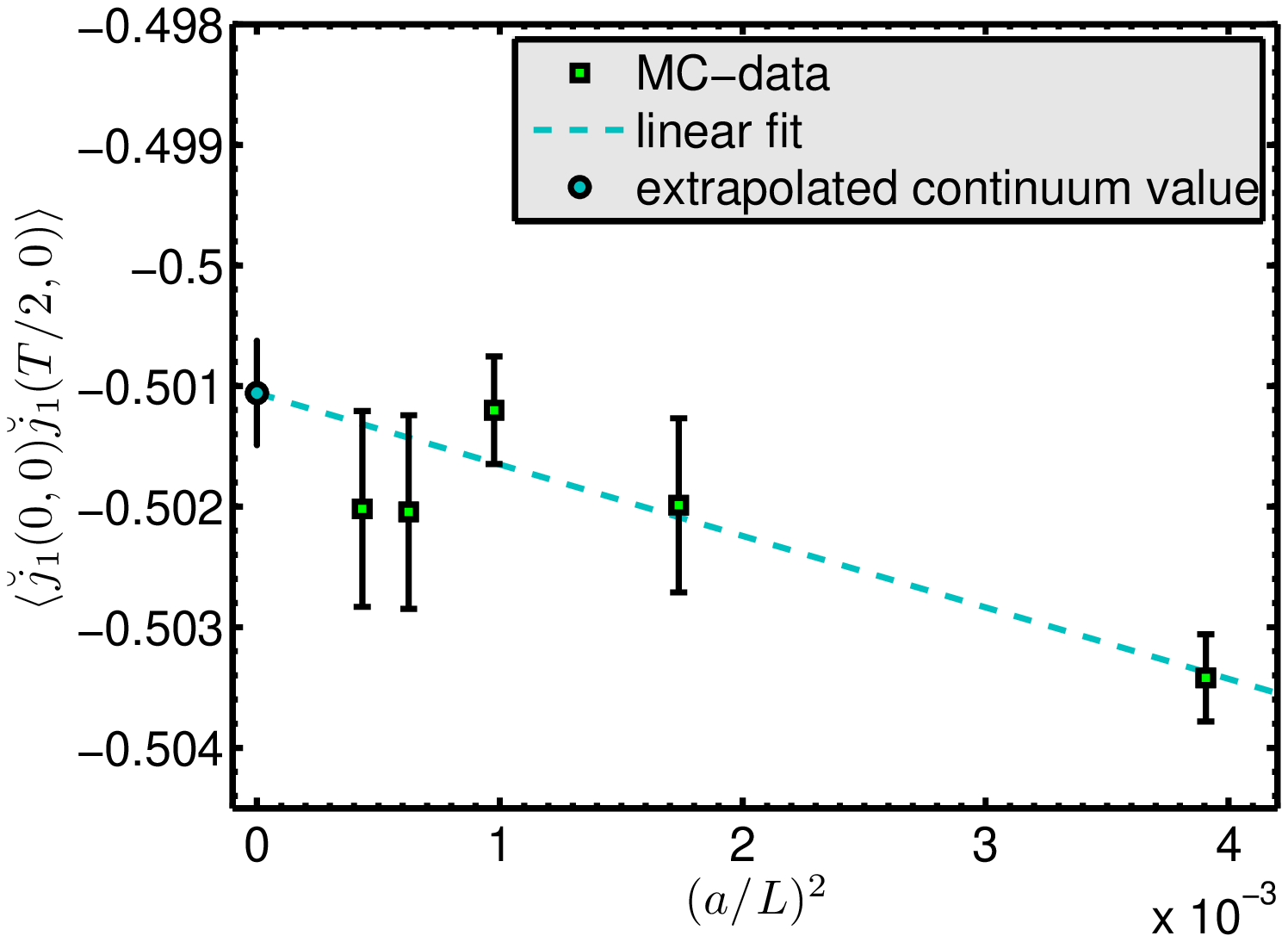}
\caption{Continuum extrapolations producing the lower point in
Fig.~\protect\ref{jjfig}.}
\label{jjextra}
\end{figure}

We have extended the well known exact continuum solution of the Thirring model
to our finite geometry and plan to report on this elsewhere \cite{TKUW}.
This allows to predict many correlation functions. As usual, correlations
of the U(1) current 
\be
\breve{j}_\mu(x_0,p_1)=\int_0^L dx_1 \, \mathe^{-ip_1 x_1} 
\bar{\psi} \gamma_\mu  \psi (x)
\ee
are particularly easy to obtain, where we use one Dirac field here at $N=2$.
In this case the symmetry (\ref{chisym}) gets automatically promoted to an axial
U(1). Correlations depend on the continuum coupling $g^{\rm cont}$.
By varying it we produce the curve in Fig.~\ref{jjfig}. Note that the quantity
on the $x$-axis is the thermal expectation value of the squared total charge
on our ring.

In the lattice transcription we employ the exact Noether U(1) current that does
not renormalize. We take the continuum limit by extrapolating from
$L/a=16\dots 48$ for values $g=0.4$ and $0.7$  of the bare
lattice coupling.  The mass is tuned to $m_c$ defined by (\ref{mccond}) on each lattice.
The extrapolations for $g=0.4$, leading to the lower point
in Fig.~\ref{jjfig} are shown in Fig.~\ref{jjextra}.
The other point is similar but with larger errors.

\section{Gross Neveu model ($N=8$)}
\begin{figure}
\centering
\includegraphics[width=.6\textwidth]{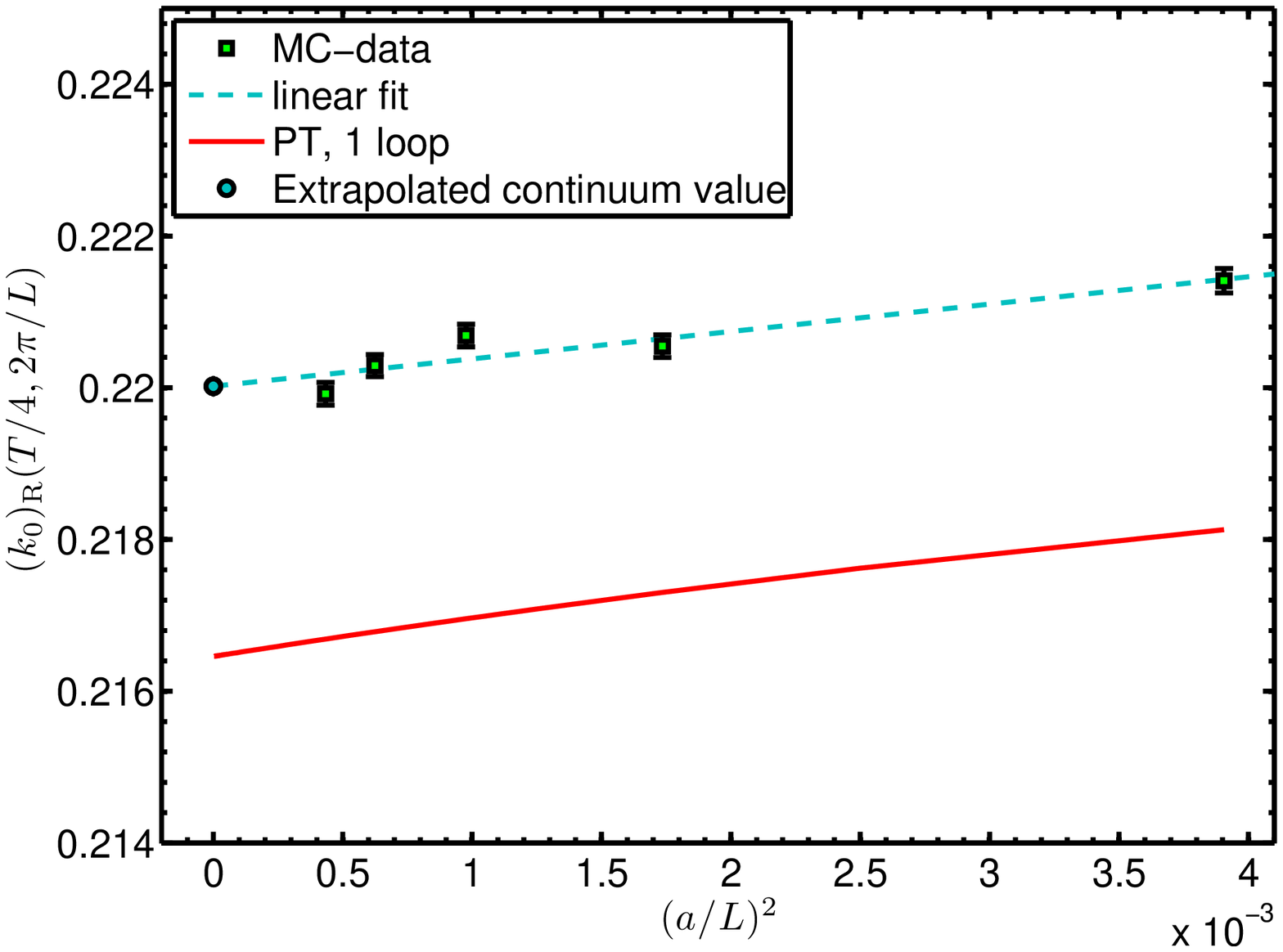}
\caption{Continuum limit of a Gross-Neveu correlation at $N=8$.}
\label{kk1fig}
\end{figure}
\begin{figure}
\centering
\includegraphics[width=.6\textwidth]{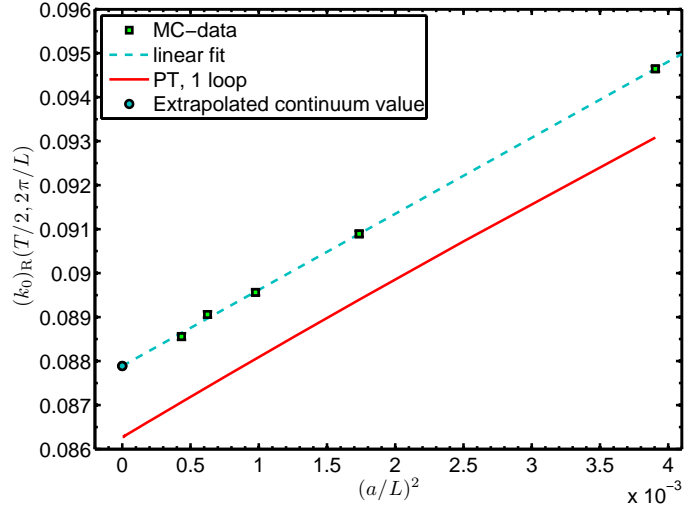}
\caption{As in Fig.~\protect\ref{kk1fig}, but for a different time separation.}
\label{kk2fig}
\end{figure}
Here the coupling renormalizes similarly to QCD.
We now adjust the coupling (\ref{gRdef}) to the value $g_R=0.38$
and take the continuum limit as discussed before.
Our universal observable in this case is
\be
(k_\mu)_R(x_0,p_1) = -\frac{1}{N Z_\xi}
a\sum_{x_1} \mathe^{-i p_1 x_1} \left\langle
\xi^\top (0) \mathcal{C} \gamma_\mu \xi(x) \right\rangle
\ee
for scalable $x_0$ values and admissible $p_1$.
Two examples are shown Fig.~\ref{kk1fig} and Fig.~\ref{kk2fig}.

\section{Remarks on staggered fermions}
For naive Majorana lattice fermions, i.~e. leaving out
the ($\propto a$) part in (\ref{Wilsonterm}), we find
the usual taste multiplicity $2^D$ in $D$ dimensions
(each momentum component around 0 or $\pi/a$). If one now
tries to `spin-diagonalize' by transforming $\xi$ one
can only achieve a reduction factor $2^{D/2-1}$
in contrast to $2^{D/2}$ for Dirac fermions when
$\psi, \bar{\psi}$ are changed independently.
The reason is that $\mathcal{C}$ can only be reduced
to $2\times 2$ blocks in the Majorana case but not
diagonalized. Since the Majorana form is natural for GN
we simply deal here with naive fermions and 4 tastes (on top
of $N$ flavors).

The 4-fermion interaction term  naively would
read in 2-momentum space
\[
\frac{1}{( T L )^4} \sum_{p_1, \ldots, p_4} \delta^2 \left( \sum p_i
   \right)  \underbrace{\tilde{\xi}_i^{\top} ( p_1 )\mathcal{C}
   \tilde{\xi}_i ( p_2 )}_{\rm taste-mixing} 
   \tilde{\xi}_j^{\top} ( p_3 )\mathcal{C} \tilde{\xi}_j ( p_4 ).
\]
As indicated, there would be contributions with for instance
$p_1$ and $p_2$ in different corners of the Brillouin zone which
mix tastes in the bilinears that are flavor-scalar.
This problem was already solved in an early effort to
simulate the model \cite{Cohen:1983nr}.
An additional factor $\prod_{\mu} \cos ( a ( p_1 + p_2 )_{\mu} / 2 )
\cos ( a ( p_3 + p_4 )_{\mu} / 2 )$ under the sum is expected to enforce
taste symmetry in the continuum limit. In position space this corresponds
to distributing the interaction term over a plaquette.
In this form one then naively expects an additional taste symmetry
in the continuum limit which may reproduce results with exact
flavor symmetry in the Wilson formulation.\\
{\bf Acknowledgement}: We would like to thank Bj\"orn Leder and Peter Weisz
for critically reading this manuscript.

\end{document}